# Bayesian Sparse Regression for Microbiome-Metabolite Data Integration

Kai Jiang[1,2], Satabdi Saha[2], Christine B. Peterson[3*]

May 18, 2026

[1]Department of Biostatistics and Data Science, The University of Texas Health Science Center at Houston, Houston, TX, USA.
[2]Department of Biostatistics, The University of Texas MD Anderson Cancer Center, Houston, TX, USA.
[3]Department of Statistics, Rice University, Houston, TX, USA.
*Correspondence to: Christine B. Peterson, cbp2@rice.edu

## Abstract

Numerous studies have shown that microbial metabolites, which represent the products of bacteria in the human gut, play a key role in shaping cancer risk and response to treatment. However, metabolite data typically contain a large proportion of missing values, which may result from either low abundance or technical challenges in data processing. Moreover, given the compositionality of microbiome data, where the observed abundances can only be interpreted on a relative scale, standard variable selection methods are not applicable. In this project, we propose a novel Bayesian regression method to address these challenges in the integration of metabolite and microbiome data. Key features of our proposed model include modeling the two different mechanisms of missingness for the metabolite data and adopting a Bayesian prior designed to address the compositional characteristics of microbiome data. We demonstrate on simulated data that our proposed model can accurately impute the unobserved true metabolite values and correctly select the relevant microbiome predictors. We further illustrate our method using real data from a study focused on understanding the interplay between the microbiome and metabolome in colorectal cancer.

# 1 Introduction

Compared to the human genome, the human microbiome contains over 150 times more genetic material [1]. The human microbiome refers to the collection of microorganisms living in and on the human body, including bacteria, viruses, and fungi [2]. The primary sites hosting the largest microbiome populations in the body are the oral cavity and the gut [3]. The gut is considered the most extensive and complex microbial environment [4].

Numerous studies have demonstrated that a balanced gut microbiome is crucial for maintaining human health. Disruption of the gut ecosystem, known as dysbiosis, has been linked to the risk of diabetes, obesity, and inflammatory bowel disease [5]. The gut microbiome can also influence cancer development by stimulating immune activity and causing chronic inflammation [6]. However, there remain a number of open questions in exactly how the microbiome impacts human disease.

Microbial metabolites are crucial to understanding the relationship between the gut microbiome and human health. Metabolites are small molecules that participate in metabolic reactions, including sugars, lipids, and amino acids. Microbial metabolites are produced or modified by microorganisms within the human body [7]. Generally, metabolites fall into three categories: 1) those produced by gut microbiota exclusively from dietary components, 2) those modified by gut microbiota but originally produced by the host, and 3) those synthesized *de novo* by the microbiota [8]. Improved modeling of the relationship between the microbiome and metabolome offers an opportunity to provide insight into how the microbiome impacts host physiology.

In characterizing the relationship between microbial features and their associated metabolites, there are some unique challenges. Microbiome data is high-dimensional, containing many microbial features, typically exceeding the number of samples. More importantly, the absolute counts for each microbial feature can vary greatly, making statistical interpretation and analysis challenging. As a result, the observed count data is typically transformed to a relative scale and treated as compositional data [9]. In this context, each row's sum equals a fixed number, usually 1 or 100%. Given that compositional data is constrained to a simplex, standard regression approaches fail to yield reliable results for variable selection and prediction of outcomes. We review existing methods for compositional regression in Section 2.

Meanwhile, metabolite data typically contain a large proportion of missing values, which are often recorded as zeros, due to low abundance or technical limitations [10]. For example, Figure 1 illustrates the distribution of thiamine, the outcome variable in our real data analysis (Section 5), after imputing zeros with half of the minimum observed value and applying a log transformation. The observed data are reasonably normally distributed, whereas the imputation of half the minimum value for zeros results in a distribution with a spike on the lower tail that clearly violates any assumptions of normality.

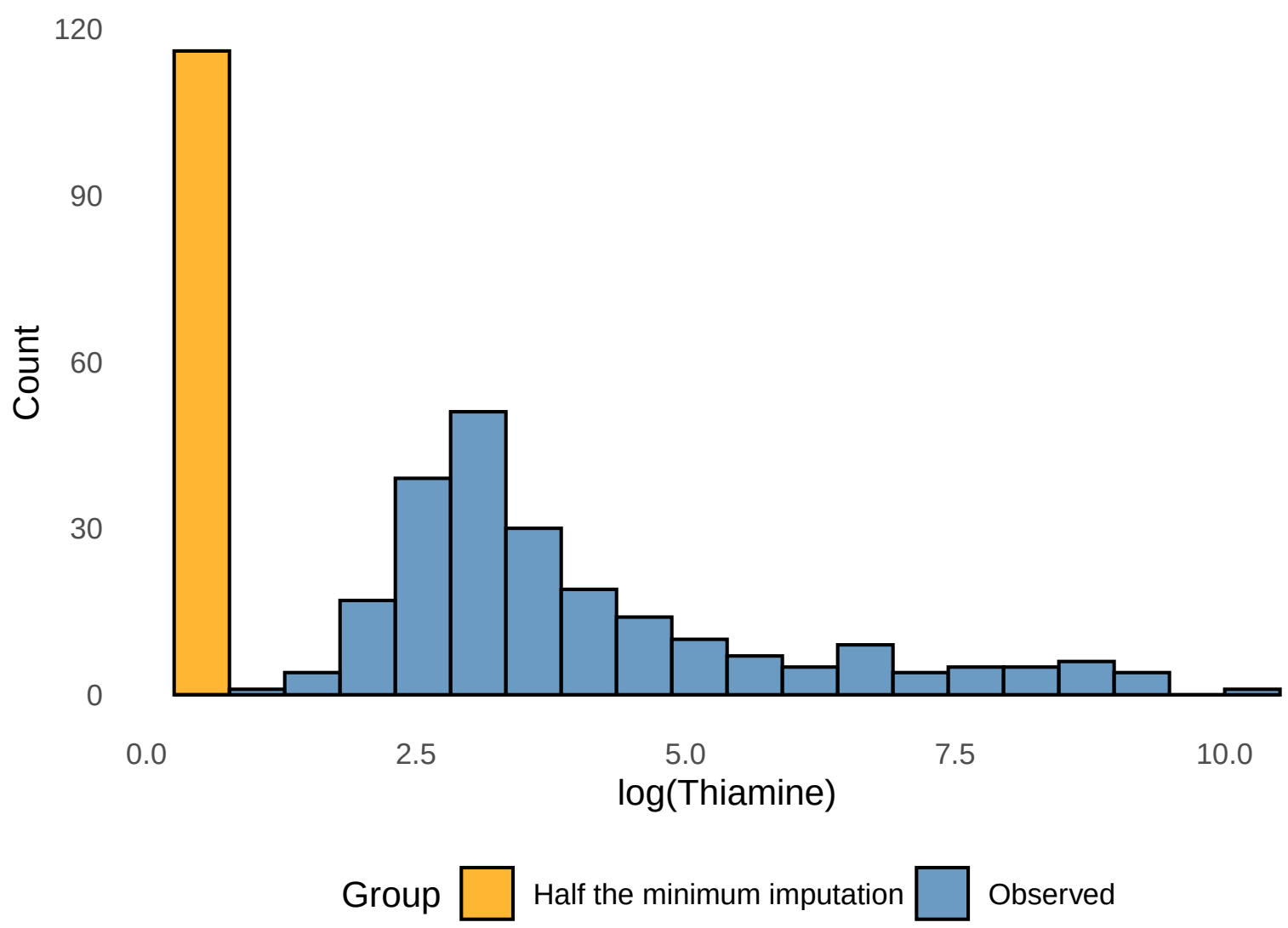

Figure 1: Distribution of thiamine abundance with zeros imputed as half the minimum value followed by log transformation.

Here, we propose a method for integrating microbiome and metabolite data, aimed at identifying microbiome features responsible for microbial metabolite production. Our proposed model, Bayesian Sparse Regression for Microbiome-Metabolite Data Integration (BSRMM), identifies a sparse set of microbiome features associated with metabolite abundance. Importantly, our model accounts for uncertainty in missing metabolite values during estimation of model parameters. Our proposed model brings together compositional regression using a $z$-prior, structured variable selection using an Ising prior, and a Bayesian truncated-normal mixture model for metabolite missingness within a unified setup. To the best of our knowledge, our proposed model represents a unique approach for seamlessly handling missing outcome values in the compositional regression framework.

The remainder of this paper is structured as follows. Section 2 provides background on missing value imputation methods and compositional data variable selection. Section 3 presents details of our proposed model. Section 4 benchmarks our model's performance in a simulation study compared with the performance of existing methods. Section 5 highlights the application of the proposed method to a study on the interplay of the microbiome and metabolites. Section 6 provides a conclusion and discussion.

# 2 Background

Missingness in metabolomic data can be broadly classified into three common sources [11]. The first occurs when the metabolite is completely absent from the patient's sample. The second arises when the metabolite's concentration is below the measurement's lower limit of detection, and the third occurs when the metabolite's concentration is above the limit of detection, but the value is missing due to technical issues.

Statistically, missing data can be categorized into three types: (i) missing completely at random (MCAR), where missingness is independent of both the observed and unobserved data; (ii) missing at random (MAR), where missingness depends only on the observed data; (iii) missing not at random (MNAR), where missingness is dependent on the unobserved data [12]. In assessing the common sources of missing data in metabolomics, Shah et al. [13] considered missing values due to technical issues as MAR, and missing values resulting from concentrations below the limit of detection as MNAR.

In practice, missing values in metabolomic data are often recorded as zeroes and are typically addressed by imputation using a fixed value, commonly half of the minimum observed concentration [10]. However, this single imputation approach is inadequate, as it assumes all missing values are MNAR and fails to account for values that may be MAR.

Moreover, as the proportion of missing values increases, the half minimum imputation becomes less effective and can substantially distort the data distribution, making it more difficult to satisfy the normality assumption, as shown in Figure 1.

As an alternative to single imputation using half of the minimum observed value, several multiple imputation methods have been developed for metabolomic data. Shah et al. [13] introduced an approach that modifies k-nearest neighbors by incorporating a truncated normal distribution. Additionally, a fully Bayesian framework has been proposed for imputing left-censored MNAR data [14]. Building on this, Shah et al. [15] proposed a Bayesian framework that simultaneously addresses both MNAR and MAR types of missingness in metabolomic data. More recently, Dekermanjian et al. [16] grouped MAR and MCAR into a combined category and proposed a two-step imputation approach that applies a random forest classifier to separately impute MNAR values and MAR/MCAR values. However, these approaches were developed for metabolite data only, with no consideration of data integration, which is the primary focus of this work.

When predicting metabolite abundances using microbiome features, missingness in the metabolite data is a key concern, as regression models typically make distributional assumptions about the response variable, but not explicitly about the predictors. However, estimation of the regression coefficients using standard methods assumes that the predictors are independent; this assumption is violated when performing regression with microbiome features as the covariates.Microbiome data are inherently compositional, meaning the features represent relative abundances that lie in a constrained simplex. Several transformation methods and statistical models have been developed to address the compositional structure of the data.

Aitchison [17] introduced the additive log-ratio (ALR) transformation, which transforms each element $u_{ij}$ of the compositional data matrix $\mathbf{U}_{n\times p}$ to $\log(u_{ij}/u_{ip})$, where $i = 1, \ldots, n$ indexes the samples and $j = 1, \ldots, p-1$ indexes the features. The denominator of the ALR transformation is a self-chosen reference variable, typically the last column from the data. Building on the ALR transformation and the linear log-contrast model of Aitchison and Bacon-Shone [18], Lin et al. [19] adopted the symmetric form of the log-contrast model and applied $l_1$ regularization to enable variable selection for compositional predictors.

Subsequently, Zhang et al. [20] developed BAZE, the first fully Bayesian approach for compositional regression with variable selection. To address compositionality in a Bayesian framework, the BAZE model introduced a generalized transformation matrix $\mathbf{T}$ within the prior structure of the regression coefficients. Combined with a spike-and-slab prior to achieve selection, the transformation approach avoids dropping reference variables and enables feature selection in the original $p$-dimensional space.

A major limitation of all the aforementioned statistical models for compositional regression is their reliance on a fully observed outcome. This requirement becomes problematic when modeling metabolite abundances as the response, as this type of data often contains a large proportion of missing values.

In summary, existing metabolomic imputation methods do not incorporate data integration and variable selection into their modeling framework. Current statistical models for compositional regression require complete responses, but metabolite outcomes often violate this assumption due to a substantial proportion of missing values. Therefore, to obtain an accurate and clear relationship between a metabolite of interest and the microbiome, we develop a novel Bayesian modeling framework that integrates both missing value imputation and variable selection simultaneously.

# 3 Methods

In this section, we describe our proposed regression model for linking a partially observed metabolite with a sparse set of microbial predictors. We begin by introducing the data structure and likelihood. We then specify priors on the latent metabolite values and regression coefficients. A unique advantage of taking a Bayesian approach to this problem is that the latent metabolite values can be considered model parameters and sampled from the posterior; this allows uncertainty about the missing values to be taken into account in estimating the remaining model parameters.

## 3.1 Data structure and likelihood

The data matrix consisting of the observed values for $p$ compositional features across $n$ samples is denoted as $\mathbf{X}_{n\times p}$, where $\mathbf{X}$ is the centered and scaled version of $\mathbf{W}$, which represents the matrix of log-transformed predictor values. We assume the vector of observed values for the metabolite of interest is represented as $\boldsymbol{m}_{n\times 1}$. For each sample $i$, where $i = 1, \ldots n$, if $m_i$ is observed, we let $y_i = \log(m_i)$, so that the metabolite values taken as our outcome variable are on a log scale. For each unobserved $m_i$, we represent the latent true value on the log scale as $y_i$.

The proposed model includes both missing value imputation on $\boldsymbol{y}$ and variable selection on $\mathbf{X}$; we will introduce the variable selection following the missing value imputation. The regression model is written as follows:

$$\boldsymbol{y} = \beta_0 \mathbf{1} + \mathbf{X}\boldsymbol{\beta} + \boldsymbol{\varepsilon}, \quad \boldsymbol{\varepsilon} \sim \mathcal{N}(\mathbf{0}, \sigma^2 \mathbf{I}), \tag{1}$$

where $\beta_0$ is the intercept term, $\mathbf{1}$ is an $n \times 1$ vector of 1s, $\boldsymbol{\beta}$ is a $p \times 1$ vector of regression coefficients, $\boldsymbol{\varepsilon}$ is an $n \times 1$ noise vector, $\sigma^2$ is a scalar variance parameter, and $\mathbf{I}$ is an $n \times n$ identity matrix. An overview figure illustrating our proposed modeling framework is provided in Figure 2 below.

## 3.2 Missing value imputation

We adopt a fully Bayesian approach to handling missing values in the regression outcome. We introduce the parameter $R_i$ to indicate whether the latent true value $y_i$ is observed. Specifically, $R_i = 1$ if $y_i$ is observed, and $R_i = 0$ if $y_i$ is missing. We model the missing data mechanism as follows:

$$P(R_i = 0 | y_i) = \begin{cases} \theta, & \text{if } y_i > \xi, \\ 1, & \text{if } y_i \leq \xi, \end{cases} \tag{2}$$

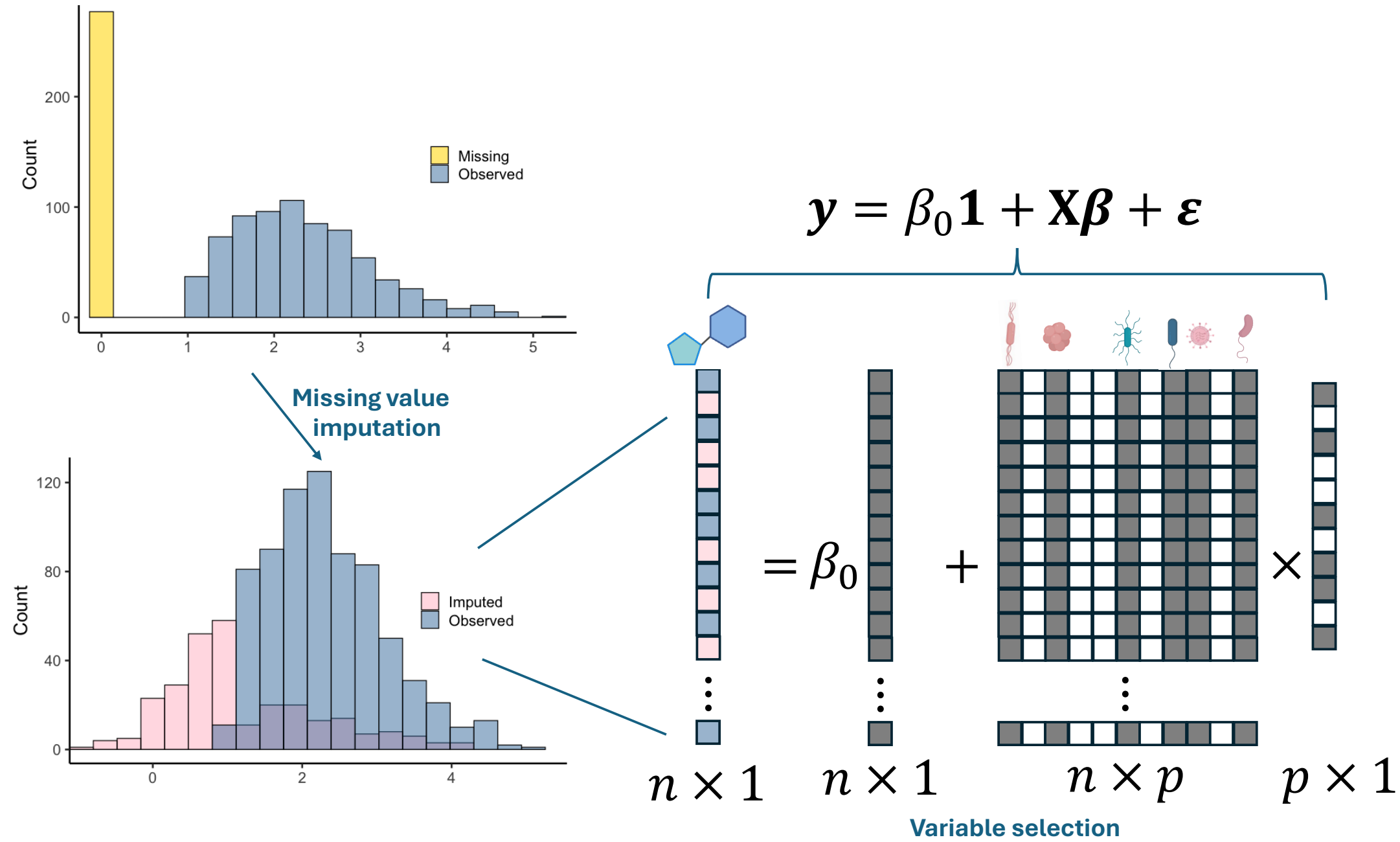

Figure 2: Overview of the proposed model. The grey color represents the selected microbial predictor variables and their corresponding nonzero coefficients.

where $\xi$ is the limit of detection, and $\theta$ denotes the probability that a latent true value above the limit of detection is still missing.

In our model, we follow the classification proposed by Shah et al. [15], which treats missingness due to low abundance as MNAR, and missingness due to technical issues as MAR. To distinguish between these two sources of missingness, we introduce a latent binary indicator $Z_i$, where $Z_i = 1$ indicates MNAR, and $Z_i = 0$ indicates MAR.

For each unobserved sample ($R_i = 0$), we assume that the latent true value $y_i$ follows a truncated normal distribution. If $Z_i = 1$, the distribution is truncated below the limit of detection; if $Z_i = 0$, the distribution is truncated above the limit of detection:

$$
\begin{aligned}
Z_i = 0\colon\ & y_i \mid R_i = 0 \sim \mathcal{N}(\beta_0 + \boldsymbol{x}_i^\top \boldsymbol{\beta}, \sigma^2) I(\xi, \infty), \\
Z_i = 1\colon\ & y_i \mid R_i = 0 \sim \mathcal{N}(\beta_0 + \boldsymbol{x}_i^\top \boldsymbol{\beta}, \sigma^2) I(-\infty, \xi).
\end{aligned} \tag{3}
$$

Let $\zeta(y_i)$ represent the normal density with mean $\beta_0 + \boldsymbol{x}_i^\top \boldsymbol{\beta}$ and variance $\sigma^2$, so that

$$\zeta(y_i) = (2\pi\sigma^2)^{-\frac{1}{2}} \exp\left\{-\frac{1}{2\sigma^2}\left(y_i - \beta_0 - \mathbf{x}_i^\top \boldsymbol{\beta}\right)^2\right\}. \tag{4}$$

Using this notation and the above framework, from equations (1), (2) and (3), we derive the probability that a missing value is missing due to being below the limit of detection as

$$P(Z_i = 1 \mid R_i = 0, \theta, \beta_0, \mathbf{x}_i, \boldsymbol{\beta}, \sigma^2) = \frac{\int_{-\infty}^{\xi} \zeta(y_i) dy_i}{\int_{-\infty}^{\xi} \zeta(y_i) dy_i + \theta \int_{\xi}^{\infty} \zeta(y_i) dy_i}. \tag{5}$$

We assume a common probability of MAR across the whole sample as $\theta$. Following Shah et al. [15], we assign a non-informative prior to $\theta$ as

$$\theta \sim \text{Unif}(0, 1). \tag{6}$$

## 3.3 Prior on the model coefficients

To handle the fixed sum constraint on compositional data, we place a $z$-prior on $\boldsymbol{\beta}$ following the work of Zhang et al. [20]. The $z$-prior is a modified version of the $g$-prior designed for high-dimensional compositional data. The formulation is motivated by the generalized lasso framework [21], which constructs a transformation matrix to accommodate different types of penalties for variable selection. In our prior, we use a latent vector of indicators $\boldsymbol{\gamma} = (\gamma_1, \ldots, \gamma_p)^\top$ to represent feature selection, where for each feature $j = 1, \ldots, p$, $\gamma_j = 1$ indicates that the $j$th feature is selected. If $\gamma_j = 0$, then the $j$th feature is omitted from the model, which is equivalent to assuming that $\beta_j = 0$. By incorporating the generalized transformation matrix

$$\mathbf{T}_{\boldsymbol{\gamma}} = \begin{bmatrix} \mathbf{I}_{p_{\boldsymbol{\gamma}}} \\ c \times \mathbf{1}_{p_{\boldsymbol{\gamma}}}^\top \end{bmatrix}_{(p_{\boldsymbol{\gamma}}+1) \times p_{\boldsymbol{\gamma}}},$$

we obtain a prior on the model coefficients of

$$\boldsymbol{\beta}_{\boldsymbol{\gamma}} \mid \sigma^2, \tau^2 \sim \mathcal{N}\left(\mathbf{0}, \sigma^2 \tau^2 (\mathbf{T}_{\boldsymbol{\gamma}}^\top \mathbf{T}_{\boldsymbol{\gamma}})^{-1}\right), \tag{7}$$

where $c$ is a constant, $\tau^2$ is a fixed hyperparameter, and $(\mathbf{T}_{\boldsymbol{\gamma}}^{\top}\mathbf{T}_{\boldsymbol{\gamma}})^{-1} = \mathbf{I}_{p_{\boldsymbol{\gamma}}} - \frac{c^2}{1+c^2 p_{\boldsymbol{\gamma}}}$. Here, $p_{\boldsymbol{\gamma}}$ represents the total number of selected features and $\boldsymbol{\beta}_{\boldsymbol{\gamma}}$ represents the vector of coefficients for the variables selected according to $\boldsymbol{\gamma}$.

Using this prior formulation, $\sum_{j\in\boldsymbol{\gamma}} \beta_j$ follows a normal distribution:

$$\sum_{j\in\boldsymbol{\gamma}} \beta_j \sim \mathcal{N}(0, \frac{p_{\boldsymbol{\gamma}}}{1+c^2 p_{\boldsymbol{\gamma}}}\sigma^2\tau^2).$$

Since the variance of $\sum_{j\in\boldsymbol{\gamma}} \beta_j$ goes to 0 when $c$ becomes large, we set $c = 100$ to enforce an approximate zero-sum constraint on the regression coefficients, following Zhang et al. [20]. Additionally, we place standard priors on the remaining parameters:

$$\sigma^2 \mid \nu, \omega \sim \text{Inverse-Gamma}\left(\frac{\nu}{2}, \frac{\nu\omega}{2}\right) \tag{8}$$

$$\beta_0 \mid \sigma^2 \sim \mathcal{N}(0, \sigma^2). \tag{9}$$

### 3.4 Ising prior for variable selection

Because our predictors of interest are microbial taxa, which often exhibit phylogenetic relatedness, accounting for this relationship between features can enhance model performance. Instead of using an independent Bernoulli prior on the inclusion probability for $\gamma_j$, we adapt the structured Ising prior framework [22, 20] to address association between microbiome features. The distribution of the Ising prior is,

$$\pi(\boldsymbol{\gamma}) = \exp(\boldsymbol{a}^T\boldsymbol{\gamma} + \boldsymbol{\gamma}^T\mathbf{Q}\boldsymbol{\gamma} - \psi(\boldsymbol{a}, \mathbf{Q})), \tag{10}$$

where $\boldsymbol{a}$ is a shrinkage parameter with $p$ elements as $\boldsymbol{a} = (a_1, \ldots, a_p)^{\top}$, $\mathbf{Q}$ is a symmetric $p \times p$ matrix of real numbers constructed from the microbiome phylogenetic tree, and $\psi(\boldsymbol{a}, \mathbf{Q})$ is the normalizing constant.

Since $\boldsymbol{\gamma}$ is binary, we can derive the conditional prior probability of including variable

$\gamma_r$, where $r \in \{1, \ldots, p\}$, from equation (10) as

$$P(\gamma_r = 1 | \boldsymbol{\gamma}_{(-r)}) = \frac{e^{\gamma_r a_r + \sum_{j \in I_{(-r)}} q_{rj} \gamma_r \gamma_j}}{1 + e^{\gamma_r a_r + \sum_{j \in I_{(-r)}} q_{rj} \gamma_r \gamma_j}}, \tag{11}$$

where $\boldsymbol{\gamma}_{(-r)}$ and $I_{(-r)}$ represent the set of indices for the remaining variables in the current model. A higher value of $q_{rj}$ in $\mathbf{Q}$ increases the prior probability that variables $r$ and $j$ are jointly selected.

### 3.5 MCMC scheme and posterior inference

The joint posterior distribution for all model parameters is:

$$p(\boldsymbol{\gamma}, \beta_0, \boldsymbol{\beta}, \sigma^2, \boldsymbol{Z}, \boldsymbol{y}, \theta \,|\, \mathbf{X}, \boldsymbol{R}) \propto p(\boldsymbol{y} \mid \beta_0, \mathbf{X}, \boldsymbol{\beta}, \boldsymbol{Z}, \boldsymbol{R}, \sigma^2) \pi(\beta_0 \mid \sigma^2) \pi(\boldsymbol{\beta} \mid \sigma^2, \boldsymbol{\gamma}) \times \\ \pi(\sigma^2) \pi(\theta \mid \boldsymbol{Z}, \boldsymbol{R}) \pi(\boldsymbol{Z} \mid \boldsymbol{R}) \pi(\boldsymbol{\gamma})$$

Since this distribution is not analytically tractable, we rely on Markov chain Monte Carlo (MCMC) to sample from the posterior. Our MCMC procedure, which follows a fully Gibbs approach, can be summarized as follows. At the $t^{\text{th}}$ iteration:

- Update the indicators of feature inclusion. Sample an index $r$ uniformly at random, and then sample the parameter $\gamma_r$ from a Bernoulli with probability:

$$P(\gamma_r = 1 | \boldsymbol{\gamma}_{(-r)}, \boldsymbol{y}, \mathbf{X}_{\boldsymbol{\gamma}}) = \frac{P(\gamma_r = 1 | \boldsymbol{\gamma}_{(-r)})}{P(\gamma_r = 1 | \boldsymbol{\gamma}_{(-r)}) + F(\boldsymbol{\gamma}' | \boldsymbol{\gamma}) \times P(\gamma_r = 0 | \boldsymbol{\gamma}_{(-r)})},$$

  where $\mathrm{F}(\gamma' \mid \boldsymbol{\gamma})^{-1}$ is the Bayes factor. Here, $\boldsymbol{\beta}$ has been integrated out to improve mixing and avoid issues with changing dimensions of the parameter space. The closed form of $\mathrm{F}(\gamma' \mid \boldsymbol{\gamma})^{-1}$ can be found in the Supplementary Materials Section S1.

- Update the regression parameters. Sample $\beta_0$, $\boldsymbol{\beta}$ for the selected variables, and $\sigma^2$:

$$\beta_0 \mid \cdot \sim \mathcal{N}\Big(\frac{\mathbf{1}^\top(\boldsymbol{y} - \mathbf{X}_\gamma \boldsymbol{\beta}_\gamma)}{n+1}, \frac{\sigma^2}{n+1}\Big),$$
$$\boldsymbol{\beta} \mid \boldsymbol{\gamma} \sim t_\nu(\hat{\boldsymbol{\beta}}_\gamma, (\mathbf{C}_\gamma + \nu\omega)\boldsymbol{a}_\gamma^{-1}),$$
$$\sigma^2 \mid \boldsymbol{\gamma} \sim \text{Inverse-Gamma}\Big(\frac{n+\nu}{2}, \frac{1}{2}[\mathbf{C}_\gamma + \nu\omega]\Big),$$

where $\mathbf{J} = \mathbf{1}^\top\mathbf{1}$, $\boldsymbol{a}_\gamma = \mathbf{X}_\gamma^\top(\mathbf{I} - \frac{\mathbf{J}}{n+1})\mathbf{X}_\gamma + \frac{\mathbf{T}_\gamma^\top\mathbf{T}_\gamma}{\tau^2}$, $\hat{\boldsymbol{\beta}}_\gamma = \boldsymbol{a}_\gamma^{-1}\mathbf{X}_\gamma^\top(\mathbf{I} - \frac{\mathbf{J}}{n+1})\boldsymbol{y}$, and $\mathbf{C}_\gamma = \boldsymbol{y}^\top(\mathbf{I} - \frac{\mathbf{J}}{n+1})\boldsymbol{y} - \boldsymbol{y}^\top(\mathbf{I} - \frac{\mathbf{J}}{n+1})\mathbf{X}_\gamma\boldsymbol{a}_\gamma^{-1}\mathbf{X}_\gamma^\top(\mathbf{I} - \frac{\mathbf{J}}{n+1})\boldsymbol{y}$.

- Update the indicator of the outcome being MNAR for observation $i$. Sample $Z_i$ for $y_i \mid R_i = 0$ from its posterior full conditional distribution, which is a Bernoulli.

- Sample the latent value $y_i \mid Z_i, R_i$ from its posterior full conditional, which is a truncated normal.

- Sample the MAR missingness probability parameter $\theta$:

$$\theta \sim \text{Beta}\Big(\sum(1 - R_i)I(y_i > \xi) + 1, \sum(R_i)I(y_i > \xi) + 1\Big).$$

After running the MCMC, we exclude the burn-in and calculate the marginal posterior probability (PPI), $p(\gamma_r \mid \boldsymbol{y})$, for each predictor across iterations. We adopt the median probability model proposed by Barbieri et al. [23] to determine the selected variables. Specifically, if $p(\gamma_r \mid \boldsymbol{y})$ exceeds 0.5, the corresponding variable is included in the final model. The details of posterior derivations are provided in the Supplementary Materials Section S1.

# 4 Simulations

In this section, we benchmark the performance of the proposed method on simulated data where the ground truth is known. This allows us to assess the accuracy of feature selection, missing value imputation, and prediction as compared to existing alternative methods.

## 4.1 Data generation

*Independent covariates*. Our simulation set-up is inspired by that of Lin et al. [19] and Zhang et al. [20]. We generate an $n \times p$ data matrix $\mathbf{O}$ from a multivariate normal distribution $\mathcal{N}(\boldsymbol{\theta}, \boldsymbol{\Sigma})$ where $n = 300$ and $p = 1000$. Because the abundance of microbiome features can vary significantly, we set the first five entries of $\boldsymbol{\theta}$ to $\log(0.5p)$ and the rest to 0. For the current simulation set-up, where we assume independent covariates, we set the covariance $\boldsymbol{\Sigma} = \mathbf{I}_p$, where $\mathbf{I}_p$ is the identity matrix. To obtain the relative abundance matrix $\mathbf{U}$, we let $u_{ij} = e^{2o_{ij}} / \sum_{k=1}^{p} e^{o_{ik}}$, where $i = 1, \ldots, 300$ and $j = 1, \ldots, 1000$.

We apply a log transformation to obtain the $\mathbf{X}$ matrix by setting $x_{ij} = \log(u_{ij})$. We set six true coefficients to nonzero values, so that our coefficient vector is given by $\boldsymbol{\beta} = \{1, -0.8, 0.6, 0, 0, -1.5, -0.5, 1.2, \mathbf{0}_{p-8}\}^T$. The signal-to-noise ratio (SNR) is calculated as $\text{mean}|\boldsymbol{\beta}_{\beta_j \neq 0}|/\sigma$. By generating different values of $\sigma$, we consider three SNR levels in our simulation study: 1, 5, and 10. Lastly, we generate the outcome variable for independent covariates from equation (1).

*Dependent covariates*. In real data applications, microbial features tend to have high levels of intercorrelation. To simulate this complexity, we construct a dependent covariate setting, where we assign 24 true coefficients as follows: $\boldsymbol{j} = \{160 + 20l\}_{l=1}^{12}$: $\boldsymbol{\beta_j} = \{0.88, -1.41, -1.39, -1.15, 1.04, 0.51, 1.21, -1.95, -1.86, 1.93, -1.34, -0.85\}$ and $\boldsymbol{j} = \{560{+}20l\}_{l=1}^{12}$: $\boldsymbol{\beta_j} = \{1.76, -1.66, -0.99, 1.48, 0.69, 1.87, -0.54, 0.72, 1.35, 0.67, -0.81, -0.16\}$. We set the covariance matrix, $\boldsymbol{\Sigma}_{ij} = 0.75 - 0.0015|i - j|$, and the mean, $\boldsymbol{\theta}_j = \log(0.5p)$, where $i, j \in \boldsymbol{j} = \{160 + 20l\}_{l=1}^{12} \cup \{560 + 20l\}_{l=1}^{12}$. This results in a set of true predictors where features that are closer together have stronger correlations.

To mimic the real data, we also allow for correlations between noise features. We set the relevant covariance matrix entries as $\boldsymbol{\Sigma}_{ij} = 0.4 - 0.02|i - j|$ and the mean as $\boldsymbol{\theta}_j = \log(0.25p)$, where $i, j \in \boldsymbol{j} = \{444 + l\}_{l=1}^{16} \cup \{944 + l\}_{l=1}^{16}$. For all remaining entries of the covariance matrix, the diagonal elements are set to be 1 and the off-diagonal elements are set to be 0.

*Missing values*. For both independent and dependent datasets, we generate missing responses following the approach detailed in Shah et al. [15]. We define three different levels of missing rates: 20%, 30%, and 40%. The missing data are generated under both MNAR and MAR mechanisms. Specifically, we consider four combinations of missing data types: (i) only MNAR, (ii) only MAR, (iii) one-third MNAR and two-thirds MAR, and (iv) two-thirds MNAR and one-third MAR. For each scenario, we randomly generate 100 datasets, with each dataset split into a 70% training set and a 30% test set.

## 4.2 Benchmarking

*Comparison methods*. We compare our model with the following existing approaches:

- Comp Lasso(i): Penalized variable selection on compositional data proposed by Lin et al. [19], with mean imputation for missing outcome values.
- Comp Lasso(ii): The same statistical method as Comp Lasso(i), but with 0.5 $\times$ the limit of detection for missing value imputation.
- BAZE(i): Bayesian framework for variable selection on compositional data proposed by Zhang et al. [20], with mean imputation for missing outcome values.
- BAZE(ii): The same statistical method as BAZE(i), but with 0.5 $\times$ the limit of detection for missing value imputation.

Adding half of the minimum observed value is the most common approach for handling missing values in metabolomics literature, but we also consider an additional single missing value imputation which assumes that the missing data are MAR.

*Model fitting*. For the Comp Lasso, we use the `Compack` package in R. Specifically, we use the `cv.compCL()` function to select the optimal hyperparameters and fit the model to the training data for variable selection. Both BAZE and our proposed model incorporate the Ising prior to capture feature similarity. Following the authors' recommendations, we set the hyperparameter $a$, which controls the overall degree of model sparsity, to $-12$ for both models.

To construct the $\mathbf{Q}$ matrix, under the independent covariates setting, we set the $\mathbf{Q}$ matrix as $\mathbf{0}_{p\times p}$. In the dependent covariates setting, where the $\mathbf{Q}$ reflects feature similarity, $\mathbf{Q}$ is structured with sparse, block-wise nonzero elements, indicating that nearby features are similar according to their phylogenetic relationship. To mimic the possibility that irrelevant features may also present similarities, we assign values to certain unrelated blocks. Accordingly, we set the elements $q_{ij} = 2.001$, where $i, j \in \{160 + 20l\}_{l=1}^{12} \cup \{560 + 20l\}_{l=1}^{12} \cup \{44 + l\}_{l=1}^{16} \cup \{444 + l\}_{l=1}^{16} \cup \{944 + l\}_{l=1}^{16}$ and $i \neq j$. The diagonal elements of $\boldsymbol{Q}$ are set to 0, and all remaining off-diagonal elements are set to 0.001.

*Performance metrics.* We assess model performance in terms of prediction, variable selection, and missing value imputation. Specifically, we evaluate the prediction error (PE) on observed samples from the test data and the L2 loss in estimation of the regression coefficient vector. Variable selection performance is assessed using the true positive rate (TPR), false positive rate (FPR), and F1 score. Finally, we evaluate the accuracy of missing value imputation on the training data using the normalized root mean squared error (NRMSE). The definitions of all evaluation metrics are provided below.

- Prediction Error:
$$\frac{\sum_{i=1}^{n^{\text{test}}} R_i (y_i^{\text{test}} - y_i^{\text{pred}})^2}{\sum_{i=1}^{n^{\text{test}}} R_i},$$
where $y_i^{\text{test}}$ is the observed outcome value for the $i$th sample in the test data.

- L2 Loss:
$$|| \hat{\boldsymbol{\beta}}_{\text{train}} - \boldsymbol{\beta}^* ||_2,$$
where $\boldsymbol{\beta}^*$ is the ground truth coefficient vector used in the simulation.

- True Positive Rate:
$$\text{TPR} = \frac{\text{TP}}{\text{TP} + \text{FN}} = \frac{\text{Number of truly non-zero coefficients correctly selected}}{\text{Total number of truly non-zero coefficients}}.$$

- False Positive Rate:

$$\text{FPR} = \frac{\text{FP}}{\text{FP} + \text{TN}} = \frac{\text{Number of truly zero coefficients incorrectly selected}}{\text{Total number of truly zero coefficients}}.$$

- F1 Score:

$$\text{F1 Score} = \frac{\text{TP}}{\text{TP} + 0.5 \times (\text{FP} + \text{FN})}.$$

- Normalized Root Mean Squared Error:

$$\sqrt{\frac{\text{mean}((\boldsymbol{y}^{\text{true}} - \boldsymbol{y}^{\text{imp}})^2)}{\text{var}(\boldsymbol{y}^{\text{true}})}},$$

  where $\boldsymbol{y}^{\text{true}}$ denotes the vector of true values, and $\boldsymbol{y}^{\text{imp}}$ represents the imputed values for unobserved samples in the training data.

## 4.3 Simulation results

In Tables 1 and 2, we present results from the independent covariate setting with 40% missing data under two scenarios: 2/3 MNAR + 1/3 MAR and 1/3 MNAR + 2/3 MAR, evaluated at three SNR levels (10, 5, and 1). BSRMM consistently outperforms the two competing methods (Comp Lasso and BAZE) across these scenarios. Specifically, BSRMM achieves the highest TPR and F1 scores, indicating its strong capability to accurately identify non-zero coefficients from high-dimensional features. Additionally, BSRMM achieves the lowest L2 Loss and NRMSE, demonstrating accuracy in both coefficient estimation and imputation under mixed MAR and MNAR missingness.

Tables 3 and 4 show results under the dependent covariate setting, which introduces a more complex covariance structure to mirror real-world data conditions. Compared to the independent setting, the performance of all three methods declines substantially due to increased noise and interdependence of the predictors. Nevertheless, BSRMM exhibits robust performance, effectively identifying non-zero variables and providing accurate estimates for coefficients and missing values. Additional simulation results and MCMC diag-nostics are available in Supplementary Materials Section S7. Figure 3 shows the imputed values from BRSMM compared to the ground-truth values for unobserved samples, where points represent posterior means and vertical bars indicate 95% credible intervals. This figure provides additional evidence for the model's ability to recover latent true values for missing samples while reflecting uncertainty in these values in the inferential framework. We present the full posterior distributions for two exemplary missing samples in Supplementary Materials Section 2.3. These two distributions, which are centered around the ground truth values, provide further evidence that our model has the capability to recover the unobserved abundance values. In summary, BSRMM consistently demonstrates strong and stable performance across various simulated scenarios.

Additionally, we evaluated the computational time of the proposed model, with detailed results reported in Supplementary Materials Section S3. The proposed model remains computationally efficient even when the number of predictors is $p = 1,000$ and the overall missing proportion is 40%, with an average runtime of approximately 70 seconds.

Patterns of MAR vs. MNAR missingness are not fully identifiable from the observed data without assumptions about the missing data mechanisms. It is therefore important to consider how our model performs under deviations from these assumptions. In particular, we assume a single fixed limit of detection $\xi$. We define this as the smallest observed non-zero value, which is a typical approach in practice. We conducted a sensitivity analysis where we varied the assumed LOD by a multiplicative scaling factor. The results indicate that the proposed model maintains stable estimation and inference across these varying LOD settings. Detailed results are provided in Supplementary Materials Section S4.

To further assess robustness, we considered three additional simulation settings. First, we considered a scenario where the MNAR mechanism was stochastic, with the probability of an observation being missing depending on its distance from the limit of detection. Second, we considered heterogeneous limits of detection across samples to reflect variability in detection thresholds. Under these two MNAR mechanisms, the proposed model demonstrates robust performance in both missing value imputation and the selection of true non-zero variables, as shown in Supplementary Materials Sections S6.1–S6.2. Finally, to evaluate sensitivity to violations of the Gaussian error assumption, outcome variables were generated using errors from a Student $t$-distribution with 3 degrees of freedom. Results in Supplementary Materials Section S6.3 indicate that the proposed model remains stable under heavy-tailed error distributions.

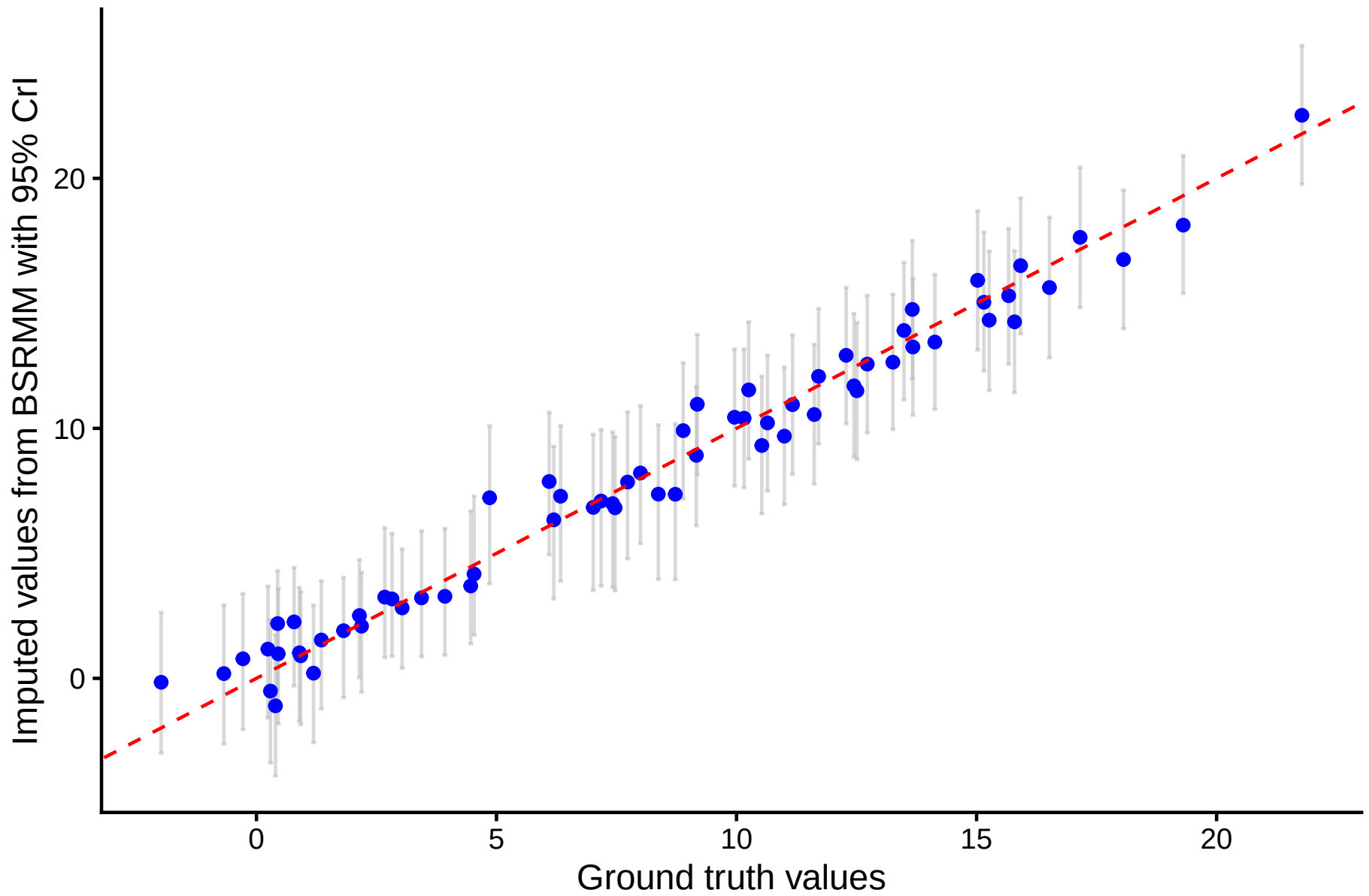

Figure 3: Comparison of imputed values from BSRMM with ground-truth values based on a simulated dataset with 30% missingness (1/3 MNAR and 2/3 MAR). Points denote posterior means, and vertical bars represent 95% credible intervals. The dashed line indicates the identity line.

Furthermore, because the $z$-prior imposes a fixed-sum constraint on the regression coefficients, we examined whether the coefficient estimates are sensitive to different proportions of missing outcomes. Our results, presented in Supplementary Materials Section S5, show that the coefficient estimates remain stable and the sum of the coefficients stays close to 0.

Table 1: Performance comparison for simulated data with independent covariate with sample size n = 300, p = 1000, missing proportion = 40% (2/3 MNAR + 1/3 MAR).

| SNR | Method | PE | L2 Loss | TPR | FPR | F1 Score | NRMSE |
|---|---|---|---|---|---|---|---|
| 10 | Comp Lasso(i) | 9.70(1.83) | 2.34(0.09) | 0.74(0.16) | 0.03(0.03) | 0.31(0.14) | 1.53(0.13) |
| | Comp Lasso(ii) | 6.32(1.87) | 1.59(0.17) | 0.98(0.06) | 0.04(0.02) | 0.29(0.11) | 1.00(0.01) |
| | BAZE(i) | 9.11(2.05) | 2.03(0.16) | 0.34(0.11) | 0.00(0.00) | 0.50(0.12) | 1.53(0.13) |
| | BAZE(ii) | 3.99(1.64) | 1.07(0.21) | 0.78(0.15) | 0.00(0.00) | 0.87(0.10) | 1.00(0.01) |
| | **BSRMM** | **0.07(0.04)** | **0.10(0.03)** | **1.00(0.00)** | **0.00(0.00)** | **1.00(0.00)** | **0.11(0.03)** |
| 5 | Comp Lasso(i) | 9.69(1.82) | 2.34(0.10) | 0.73(0.18) | 0.03(0.03) | 0.29(0.15) | 1.53(0.14) |
| | Comp Lasso(ii) | 6.50(1.84) | 1.58(0.17) | 0.96(0.08) | 0.04(0.02) | 0.28(0.11) | 1.00(0.01) |
| | BAZE(i) | 9.12(2.11) | 2.03(0.17) | 0.34(0.11) | 0.00(0.00) | 0.49(0.13) | 1.53(0.14) |
| | BAZE(ii) | 4.00(1.72) | 1.06(0.24) | 0.79(0.17) | 0.00(0.00) | 0.87(0.12) | 1.00(0.01) |
| | **BSRMM** | **0.09(0.04)** | **0.10(0.03)** | **1.00(0.00)** | **0.00(0.00)** | **1.00(0.00)** | **0.11(0.03)** |
| 1 | Comp Lasso(i) | 10.17(1.90) | 2.35(0.09) | 0.71(0.17) | 0.03(0.02) | 0.30(0.13) | 1.53(0.12) |
| | Comp Lasso(ii) | 7.48(2.15) | 1.62(0.17) | 0.96(0.10) | 0.04(0.02) | 0.29(0.13) | 1.00(0.01) |
| | BAZE(i) | 9.78(2.27) | 2.06(0.22) | 0.33(0.15) | 0.00(0.00) | 0.47(0.18) | 1.53(0.12) |
| | BAZE(ii) | 5.01(1.75) | 1.11(0.20) | 0.74(0.16) | 0.00(0.00) | 0.84(0.11) | 1.00(0.01) |
| | **BSRMM** | **0.99(0.27)** | **0.14(0.06)** | **1.00(0.02)** | **0.00(0.00)** | **1.00(0.01)** | **0.22(0.04)** |

Table 2: Performance comparison for simulated data with independent covariate with sample size n = 300, p = 1000, missing proportion = 40% (1/3 MNAR + 2/3 MAR).

| SNR | Method | PE | L2 Loss | TPR | FPR | F1 Score | NRMSE |
|---|---|---|---|---|---|---|---|
| 10 | Comp Lasso(i) | 10.53(1.95) | 2.15(0.11) | 0.84(0.13) | 0.03(0.03) | 0.29(0.13) | 1.15(0.06) |
| | Comp Lasso(ii) | 14.33(4.05) | 1.89(0.20) | 0.85(0.14) | 0.03(0.03) | 0.29(0.13) | 1.26(0.06) |
| | BAZE(i) | 9.29(2.63) | 1.79(0.20) | 0.46(0.15) | 0.00(0.00) | 0.61(0.14) | 1.15(0.06) |
| | BAZE(ii) | 12.25(4.25) | 1.45(0.29) | 0.49(0.15) | 0.00(0.00) | 0.64(0.13) | 1.26(0.06) |
| | **BSRMM** | **0.09(0.23)** | **0.09(0.08)** | **1.00(0.02)** | **0.00(0.00)** | **1.00(0.01)** | **0.08(0.06)** |
| 5 | Comp Lasso(i) | 10.43(1.80) | 2.15(0.12) | 0.83(0.13) | 0.03(0.02) | 0.29(0.13) | 1.15(0.06) |
| | Comp Lasso(ii) | 14.46(4.06) | 1.88(0.19) | 0.85(0.14) | 0.04(0.03) | 0.28(0.12) | 1.26(0.05) |
| | BAZE(i) | 9.06(2.12) | 1.78(0.17) | 0.47(0.14) | 0.00(0.00) | 0.62(0.12) | 1.15(0.06) |
| | BAZE(ii) | 12.03(3.69) | 1.44(0.28) | 0.50(0.15) | 0.00(0.00) | 0.65(0.14) | 1.26(0.05) |
| | **BSRMM** | **0.69(3.54)** | **0.16(0.36)** | **0.97(0.15)** | **0.00(0.00)** | **0.97(0.14)** | **0.11(0.15)** |
| 1 | Comp Lasso(i) | 11.32(2.16) | 2.17(0.12) | 0.82(0.14) | 0.03(0.03) | 0.29(0.14) | 1.14(0.06) |
| | Comp Lasso(ii) | 15.70(4.58) | 1.94(0.20) | 0.83(0.12) | 0.03(0.02) | 0.29(0.14) | 1.26(0.05) |
| | BAZE(i) | 9.79(2.37) | 1.80(0.18) | 0.46(0.15) | 0.00(0.00) | 0.61(0.15) | 1.14(0.06) |
| | BAZE(ii) | 13.57(4.73) | 1.51(0.32) | 0.46(0.16) | 0.00(0.00) | 0.61(0.16) | 1.26(0.05) |
| | **BSRMM** | **1.63(3.60)** | **0.28(1.02)** | **0.97(0.16)** | **0.00(0.00)** | **0.97(0.15)** | **0.21(0.11)** |

Table 3: Performance comparison for simulated data with dependent covariate with sample size n = 300, p = 1000, missing proportion = 40% (2/3 MNAR + 1/3 MAR).

| SNR | Method | PE | L2 Loss | TPR | FPR | F1 Score | NRMSE |
|---|---|---|---|---|---|---|---|
| 10 | Comp Lasso(i) | 54.50(9.59) | 5.99(0.13) | 0.32(0.07) | 0.02(0.02) | 0.33(0.09) | 3.61(0.39) |
| | Comp Lasso(ii) | 47.65(14.68) | 5.57(0.15) | 0.40(0.07) | 0.02(0.02) | 0.38(0.11) | 1.00(0.01) |
| | BAZE(i) | 52.64(9.04) | 5.84(0.34) | 0.17(0.19) | 0.02(0.00) | 0.17(0.16) | 3.61(0.39) |
| | BAZE(ii) | 31.06(17.39) | 4.05(1.15) | 0.70(0.34) | 0.01(0.01) | 0.59(0.25) | 1.00(0.01) |
| | **BSRMM** | **0.10(0.04)** | **0.23(0.04)** | **1.00(0.00)** | **0.01(0.01)** | **0.82(0.11)** | **0.03(0.01)** |
| 5 | Comp Lasso(i) | 53.62(9.57) | 5.99(0.12) | 0.33(0.08) | 0.02(0.02) | 0.33(0.10) | 3.66(0.40) |
| | Comp Lasso(ii) | 47.37(14.46) | 5.57(0.14) | 0.42(0.07) | 0.03(0.02) | 0.37(0.11) | 1.00(0.01) |
| | BAZE(i) | 53.08(9.77) | 5.88(0.34) | 0.15(0.18) | 0.02(0.00) | 0.16(0.15) | 3.66(0.40) |
| | BAZE(ii) | 33.45(17.77) | 4.22(1.13) | 0.67(0.33) | 0.02(0.00) | 0.56(0.24) | 1.00(0.01) |
| | **BSRMM** | **0.17(0.09)** | **0.26(0.08)** | **1.00(0.00)** | **0.01(0.01)** | **0.81(0.11)** | **0.04(0.01)** |
| 1 | Comp Lasso(i) | 54.88(9.82) | 6.00(0.11) | 0.32(0.08) | 0.02(0.02) | 0.33(0.11) | 3.56(0.40) |
| | Comp Lasso(ii) | 47.83(12.67) | 5.58(0.14) | 0.40(0.06) | 0.02(0.02) | 0.38(0.11) | 1.00(0.01) |
| | BAZE(i) | 54.29(9.10) | 5.92(0.31) | 0.12(0.14) | 0.02(0.00) | 0.13(0.11) | 3.56(0.40) |
| | BAZE(ii) | 32.04(15.98) | 4.08(1.10) | 0.70(0.33) | 0.02(0.00) | 0.57(0.24) | 1.00(0.01) |
| | **BSRMM** | **1.91(0.41)** | **0.59(0.09)** | **1.00(0.00)** | **0.01(0.01)** | **0.81(0.11)** | **0.12(0.01)** |

Table 4: Performance comparison for simulated data with dependent covariate with sample size n = 300, p = 1000, missing proportion = 40% (1/3 MNAR + 2/3 MAR).

| SNR | Method | PE | L2 Loss | TPR | FPR | F1 Score | NRMSE |
|---|---|---|---|---|---|---|---|
| 10 | Comp Lasso(i) | 65.25(15.10) | 5.98(0.12) | 0.32(0.08) | 0.02(0.02) | 0.32(0.09) | 1.14(0.06) |
| | Comp Lasso(ii) | 87.10(24.65) | 5.79(0.14) | 0.34(0.08) | 0.02(0.02) | 0.34(0.11) | 1.22(0.05) |
| | BAZE(i) | 60.56(14.36) | 5.80(0.41) | 0.20(0.21) | 0.02(0.00) | 0.20(0.18) | 1.14(0.06) |
| | BAZE(ii) | 83.56(22.09) | 5.87(0.64) | 0.20(0.23) | 0.01(0.01) | 0.20(0.19) | 1.22(0.05) |
| | **BSRMM** | **0.10(0.06)** | **0.24(0.06)** | **1.00(0.00)** | **0.01(0.01)** | **0.80(0.10)** | **0.03(0.01)** |
| 5 | Comp Lasso(i) | 64.61(13.94) | 5.96(0.12) | 0.33(0.07) | 0.02(0.02) | 0.33(0.09) | 1.14(0.05) |
| | Comp Lasso(ii) | 87.37(24.02) | 5.79(0.14) | 0.33(0.08) | 0.02(0.03) | 0.32(0.11) | 1.22(0.05) |
| | BAZE(i) | 60.30(13.44) | 5.79(0.45) | 0.19(0.20) | 0.01(0.00) | 0.19(0.18) | 1.14(0.05) |
| | BAZE(ii) | 83.38(20.54) | 5.93(0.58) | 0.16(0.19) | 0.01(0.01) | 0.18(0.17) | 1.22(0.05) |
| | **BSRMM** | **0.16(0.06)** | **0.26(0.05)** | **1.00(0.00)** | **0.01(0.01)** | **0.80(0.10)** | **0.03(0.01)** |
| 1 | Comp Lasso(i) | 65.15(13.75) | 5.97(0.12) | 0.33(0.07) | 0.02(0.02) | 0.34(0.09) | 1.15(0.06) |
| | Comp Lasso(ii) | 84.60(21.86) | 5.78(0.15) | 0.34(0.07) | 0.02(0.02) | 0.33(0.09) | 1.22(0.05) |
| | BAZE(i) | 60.76(14.55) | 5.82(0.35) | 0.17(0.19) | 0.02(0.00) | 0.17(0.16) | 1.15(0.06) |
| | BAZE(ii) | 81.32(19.21) | 5.80(0.71) | 0.25(0.26) | 0.01(0.00) | 0.24(0.22) | 1.22(0.05) |
| | **BSRMM** | **1.93(0.43)** | **0.59(0.11)** | **1.00(0.00)** | **0.01(0.01)** | **0.82(0.11)** | **0.10(0.01)** |

# 5 Application to colorectal cancer data

In our case study, we seek to characterize microbial features that shape the abundance of thiamine in the gut. We focus on thiamine, also known as vitamin B1, as meta-analyses have shown that higher thiamine levels are associated with a reduced risk of colorectal cancer [24]. The primary source of thiamine for the human body is dietary intake, including whole grains, meat, and fish [25]. Although the human body cannot synthesize thiamine endogenously, certain gut microbiota are considered a potential source for thiamine biosynthesis [26, 27].

To carry out this analysis, we relied on a data set originally obtained by Yachida et al. [28] and subsequently included in The Curated Gut Microbiome Metabolome Data Resource [29]. This study included both microbiome and metabolic profiling on 347 subjects. Metabolite concentrations were measured by capillary electrophoresis time-of-flight mass spectrometry, and microbiome features data were quantified from shotgun metagenomics sequencing. As shown in Figure 1, thiamine exhibits a high proportion of missing values in this data set, approximately 35%.

To obtain a suitable set of microbial predictors, we filtered the available species on both prevalence and abundance, focusing on features with prevalence of at least 30% and average relative abundance of at least 0.3%. After filtering, we obtained 407 microbial features for inclusion as candidate predictors in our model. For relative abundances equal to zero, we added a pseudocount of 0.5 $\times$ the minimum observed value to allow log transformation [30].

We applied all methods considered in the simulation study in order to compare model sparsity and prediction error. In applying our proposed method, we sought to construct a $\mathbf{Q}$ matrix so that features with closer phylogenetic relatedness would have more strongly positive entries to encourage the joint selection of similar features. To achieve this, we computed the partial phylogenetic correlation matrix, and replaced all diagonal and negative elements with zeros. Since the shrinkage parameter $\boldsymbol{a}$ in the Ising prior controls predictor sparsity, we performed a sensitivity analysis on $\boldsymbol{a}$ for BSRMM, BAZE(i), and BAZE(ii).

When $\boldsymbol{a} = \mathbf{-6}$, all three models selected a comparable and reasonable number of features. Increasing $\boldsymbol{a}$ beyond $\mathbf{-6}$ resulted in overly dense models. Accordingly, we set $\boldsymbol{a} = \mathbf{-6}$. All other hyperparameters were consistent with those used in the simulation study. Additional details on the partial phylogenetic correlation matrix $\mathbf{Q}$ and the sensitivity analysis for $\boldsymbol{a}$ are provided in the Supplementary Materials Section Section S8 and S9. For Comp Lasso (i) and Comp Lasso (ii), we applied the same model fitting procedures as in the simulation settings.

To compare the methods' prediction, we randomly split the dataset into a 70% training set and a 30% test set. As in the simulation study, we considered features with PPI $> 0.5$ as selected for both BAZE and BSRMM. Table 5 presents the number of selected variables and prediction error on the test set for the proposed model and comparison methods. BSRMM achieves the lowest prediction error, identifying six microbial features associated with thiamine levels belonging to three unique families: *Bacteroidaceae*, *Enterobacteriaceae*, and *Lachnospiraceae*, as shown in Table 6. The alternative methods considered all selected fewer features. Although the ground truth is not known, this may suggest that BSRMM offers improved power over existing approaches, consistent with the higher TPR for BSRMM observed in the simulation study.

Among the features selected by BSRMM, four were from the family *Lachnospiraceae*, one was from *Bacteroidaceae*, and one was from *Enterobacteriaceae*. Importantly, vitamin biosynthesis in the gut relies on multiple interacting bacteria [26]. Recent research has shown that most strains belonging to *Lachnospiraceae* have the potential to produce thiamine, although some strains may rely on it as an intake from their environment [27], supporting our finding of both negative and positive associations within this family. In addition, *Prevotella* has also been associated with B-vitamin synthesis [31]. Our findings characterize the connection between microbiota and thiamine abundance in the human gut, highlighting a potential mechanism for how the gut microbiome may influence colorectal cancer risk [32].

Table 5: Model performance summary for real data analysis.

| Model | Number of Selected Variables | Prediction Error |
|---|---|---|
| Comp Lasso (i) | 3 | 4.63 |
| Comp Lasso (ii) | 0 | 6.26 |
| BAZE (i) | 5 | 4.89 |
| BAZE (ii) | 2 | 9.39 |
| **BSRMM** | **6** | **4.47** |

Table 6: Microbial features selected by the BSRMM model with 95% credible intervals (CrI) for thiamine outcome.

| Family | Genus | Species | Lower 95% CrI | Posterior Mean | Upper 95% CrI |
|---|---|---|---|---|---|
| Bacteroidaceae | Prevotella | Prevotella_sp900770515 | -1.35 | -0.63 | -0.04 |
| Enterobacteriaceae | Escherichia | Escherichia_sp001660175 | -2.69 | -0.88 | 0.01 |
| Lachnospiraceae | Anaerostipes | Anaerostipes_sp905215045 | -1.28 | -0.56 | -0.08 |
| Lachnospiraceae | Enterocloster | Enterocloster_citroniae | -1.13 | -0.56 | -0.04 |
| Lachnospiraceae | Enterocloster | Enterocloster_sp001517625 | 0.07 | 0.49 | 0.92 |
| Lachnospiraceae | Hungatella | Hungatella_sp005845265 | 0.21 | 0.61 | 0.99 |

# 6 Discussion

In this project, we developed a fully Bayesian framework to investigate the relationship between microbiome features and metabolite levels. Our proposed model addresses three major challenges: the high-dimensionality and compositional nature of microbiome data and the presence of missing values in the metabolite outcome variable. Our proposed BSRMM model enables accurate variable selection by identifying non-zero coefficients from a large set of predictors while simultaneously estimating the latent true values of partially observed metabolites.

This study has two main limitations. First, for the real dataset, the true values of the missing outcomes are unknown. As a result, we are unable to directly evaluate the performance of our missing value imputation on the real data. Second, the current model framework includes only a single outcome. However, in metabolomics research, metabolites are often part of known metabolic pathways, where multiple compounds are connected through biochemical reactions and act together. In future work, we plan to extend this framework to a multivariate outcome setting to better capture the underlying biological mechanisms.

# 7 Acknowledgment

The authors thank Dr. Liangliang Zhang and Mr. Ruitao Liu for providing the R code used in the paper for BAZE model.

# 8 Funding

This research was partially funded by NIH R01 HL158796, NIH/NCI CCSG P30CA016672, and an Andrew Sabin Family Fellowship.

# 9 Data and Code Availability

The metabolite and microbiome data [28] analyzed in Section 5 was curated by Muller et al. [29] and is available online at https://github.com/borenstein-lab/microbiome-metabolome-curated-data/wiki.

The implementation of the proposed BSRMM model is available at https://github.com/Kai-Jiang-1/BSRMM.

# References

[1] Kaijian Hou, Zhuo-Xun Wu, Xuan-Yu Chen, Jing-Quan Wang, Dongya Zhang, Chuanxing Xiao, Dan Zhu, Jagadish B Koya, Liuya Wei, Jilin Li, et al. Microbiota in health and diseases. *Signal Transduction and Targeted Therapy*, 7(1):1–28, 2022.

[2] Emanuele Rinninella, Pauline Raoul, Marco Cintoni, Francesco Franceschi, Giacinto Abele Donato Miggiano, Antonio Gasbarrini, and Maria Cristina Mele. What is the healthy gut microbiota composition? A changing ecosystem across age, environment, diet, and diseases. *Microorganisms*, 7(1):14, 2019.

[3] Luke K Ursell, Henry J Haiser, Will Van Treuren, Neha Garg, Lavanya Reddivari, Jairam Vanamala, Pieter C Dorrestein, Peter J Turnbaugh, and Rob Knight. The intestinal metabolome: an intersection between microbiota and host. *Gastroenterology*, 146(6):1470–1476, 2014.

[4] Sigal Leviatan, Saar Shoer, Daphna Rothschild, Maria Gorodetski, and Eran Segal. An expanded reference map of the human gut microbiome reveals hundreds of previously unknown species. *Nature Communications*, 13(1):3863, 2022.

[5] Gillian M Barlow, Allen Yu, and Ruchi Mathur. Role of the gut microbiome in obesity and diabetes mellitus. *Nutrition in Clinical Practice*, 30(6):787–797, 2015.

[6] Noor Akbar, Naveed Ahmed Khan, Jibran Sualeh Muhammad, and Ruqaiyyah Siddiqui. The role of gut microbiome in cancer genesis and cancer prevention. *Health Sciences Review*, 2:100010, 2022.

[7] Jong-Hwi Yoon, Jun-Soo Do, Priyanka Velankanni, Choong-Gu Lee, and Ho-Keun Kwon. Gut microbial metabolites on host immune responses in health and disease. *Immune Network*, 23(1):e6, 2023.

[8] Panida Sittipo, Jae-won Shim, and Yun Kyung Lee. Microbial metabolites determine host health and the status of some diseases. *International Journal of Molecular Sciences*, 20(21):5296, 2019.

[9] Hongzhe Li. Microbiome, metagenomics, and high-dimensional compositional data analysis. *Annual Review of Statistics and Its Application*, 2:73–94, 2015.

[10] Trenton J Davis, Tarek R Firzli, Emily A Higgins Keppler, Matthew Richardson, and Heather D Bean. Addressing missing data in GC × GC metabolomics: Identifying missingness type and evaluating the impact of imputation methods on experimental replication. *Analytical Chemistry*, 94(31):10912–10920, 2022.

[11] Sandra L Taylor, Gary S Leiserowitz, and Kyoungmi Kim. Accounting for undetected compounds in statistical analyses of mass spectrometry ‘omic studies. *Statistical Applications in Genetics and Molecular Biology*, 12(6):703–722, 2013.

[12] Donald B Rubin. *Statistical Analysis with Missing Data.* Wiley, 1987.

[13] Jasmit S Shah, Shesh N Rai, Andrew P DeFilippis, Bradford G Hill, Aruni Bhatnagar, and Guy N Brock. Distribution based nearest neighbor imputation for truncated high dimensional data with applications to pre-clinical and clinical metabolomics studies. *BMC Bioinformatics*, 18:1–13, 2017.

[14] Runmin Wei, Jingye Wang, Erik Jia, Tianlu Chen, Yan Ni, and Wei Jia. Gsimp: A gibbs sampler based left-censored missing value imputation approach for metabolomics studies. *PLoS Computational Biology*, 14(1):e1005973, 2018.

[15] Jasmit Shah, Guy N Brock, and Jeremy Gaskins. BayesMetab: Treatment of missing values in metabolomic studies using a bayesian modeling approach. *BMC Bioinformatics*, 20(Suppl 24):673, 2019.

[16] Jonathan P Dekermanjian, Elin Shaddox, Debmalya Nandy, Debashis Ghosh, and Katerina Kechris. Mechanism-aware imputation: a two-step approach in handling missing values in metabolomics. *BMC Bioinformatics*, 23(1):179, 2022.

[17] John Aitchison. The statistical analysis of compositional data. *Journal of the Royal Statistical Society: Series B (Methodological)*, 44(2):139–160, 1982.

[18] John Aitchison and John Bacon-Shone. Log contrast models for experiments with mixtures. *Biometrika*, 71(2):323–330, 1984.

[19] Wei Lin, Pixu Shi, Rui Feng, and Hongzhe Li. Variable selection in regression with compositional covariates. *Biometrika*, 101(4):785–797, 2014.

[20] Liangliang Zhang, Yushu Shi, Robert R Jenq, Kim-Anh Do, and Christine B Peterson. Bayesian compositional regression with structured priors for microbiome feature selection. *Biometrics*, 77(3):824–838, 2021.

[21] Ryan J Tibshirani and Jonathan Taylor. The solution path of the generalized lasso. *The Annals of Statistics*, 39(3):1335, 2011.

[22] Fan Li and Nancy R Zhang. Bayesian variable selection in structured high-dimensional covariate spaces with applications in genomics. *Journal of the American Statistical Association*, 105(491):1202–1214, 2010.

[23] Maria M Barbieri, James O Berger, Edward I George, and Veronika Ročková. The median probability model and correlated variables. *Bayesian Analysis*, 16(4):1085–1112, 2021.

[24] Yan Liu, Wen-jing Xiong, Lei Wang, Chuanhua YU, et al. Vitamin B1 intake and the risk of colorectal cancer: a systematic review of observational studies. *Journal of Nutritional Science and Vitaminology*, 67(6):391–396, 2021.

[25] John W. Erdman, Ian A. MacDonald, and Steven H. Zeisel. *Present Knowledge in Nutrition: Tenth Edition*. Wiley-Blackwell, United States, June 2012. ISBN 9780470959176. doi: 10.1002/9781119946045.

[26] Stefanía Magnúsdóttir, Dmitry Ravcheev, Valérie de Crécy-Lagard, and Ines Thiele. Systematic genome assessment of b-vitamin biosynthesis suggests co-operation among gut microbes. *Frontiers in Genetics*, 6:148, 2015.

[27] Chiara Tarracchini, Gabriele Andrea Lugli, Leonardo Mancabelli, Douwe van Sinderen, Francesca Turroni, Marco Ventura, and Christian Milani. Exploring the vitamin biosynthesis landscape of the human gut microbiota. *mSystems*, 9(10):e00929–24, 2024.

[28] Shinichi Yachida, Sayaka Mizutani, Hirotsugu Shiroma, Satoshi Shiba, Takeshi Nakajima, Taku Sakamoto, Hikaru Watanabe, Keigo Masuda, Yuichiro Nishimoto, Masaru Kubo, et al. Metagenomic and metabolomic analyses reveal distinct stage-specific phenotypes of the gut microbiota in colorectal cancer. *Nature Medicine*, 25(6):968–976, 2019.

[29] Efrat Muller, Yadid M Algavi, and Elhanan Borenstein. The gut microbiome-metabolome dataset collection: a curated resource for integrative meta-analysis. *npj Biofilms and Microbiomes*, 8(1):79, 2022.

[30] Yoshihiko Tomofuji, Toshihiro Kishikawa, Kyuto Sonehara, Yuichi Maeda, Kotaro Ogawa, Shuhei Kawabata, Eri Oguro-Igashira, Tatsusada Okuno, Takuro Nii, Makoto Kinoshita, et al. Analysis of gut microbiome, host genetics, and plasma metabolites reveals gut microbiome-host interactions in the japanese population. *Cell Reports*, 42 (11), 2023.

[31] IN Abdurasulova, EA Chernyavskaya, AB Ivanov, VA Nikitina, VI Lioudyno, AA Nartova, AV Matsulevich, E Yu Skripchenko, GN Bisaga, VI Ulyantsev, et al. Changes in gut microbiome taxonomic composition and their relationship to biosynthetic and metabolic pathways of b vitamins in children with multiple sclerosis. *Journal of Evolutionary Biochemistry and Physiology*, 60(1):114–135, 2024.

[32] Minsuk Kim, Emily Vogtmann, David A Ahlquist, Mary E Devens, John B Kisiel, William R Taylor, Bryan A White, Vanessa L Hale, Jaeyun Sung, Nicholas Chia, et al. Fecal metabolomic signatures in colorectal adenoma patients are associated with gut microbiota and early events of colorectal cancer pathogenesis. *MBio*, 11(1): 10–1128, 2020.